# Experimental Study of the Exciton Gas – Liquid Transition in Coupled Quantum Wells

Subhradeep Misra[1], Michael Stern[2], Arjun Joshua[1], Vladimir Umansky[1] and Israel Bar-Joseph[1]

[1]*Department of Condensed Matter physics, Weizmann Institute of Science, Rehovot, Israel*
[2]*Institute of Nanotechnology and Advanced Materials, Bar-Ilan University, Ramat-Gan, Israel*

**Abstract**

*We study the exciton gas-liquid transition in GaAs/AlGaAs coupled quantum wells. Below a critical temperature, $T_C=4.8K$, and above a threshold laser power density the system undergoes a phase transition into a liquid state. We determine the density – temperature phase diagram over the temperature range 0.1 – 4.8K. We find that the latent heat increases linearly with temperature at $T \lesssim 1.1K$, similarly to a Bose Einstein condensate transition, and becomes constant at $1.1 \lesssim T < 4.8K$. Resonant Rayleigh scattering measurements reveal that the disorder in the sample is strongly suppressed and the diffusion coefficient sharply increases with decreasing temperature at $T<T_C$, allowing the liquid to spread over large distances away from the excitation region. We suggest that our findings are manifestations of a quantum liquid behavior.*

Dipolar excitons in coupled quantum wells (CQW) offer an interesting test bed for studying collective effects of an interacting quantum degenerate system [1, 2]. Their relatively light mass, which is smaller than that of a free electron, implies that the necessary conditions for achieving quantum degeneracy can occur already at cryogenic temperatures, and their strong dipole-dipole interaction may give rise to formation of ordered phases. Extensive attempts have been made over the past two decades to observe these phases and to determine the phase diagram of this system [3- 12].

In recent years there are mounting evidences for a phase transition that occurs at low temperatures in this system, yet its nature and thermodynamics remain open questions. Many of the studies are performed in a trap geometry, which confines the excitons to a narrow region around the illuminated spot [13, 14], and evidences for condensation are found through photoluminescence (PL) anomalies: The appearance of spontaneous coherence [7], onset of non-radiative recombination ("PL darkening") [9] and large blueshift of the PL energy [10, 11]. An alternative approach to study this phase transition uses an open geometry, where photo-excited carriers are free to move away from the illumination spot, and their diffusion is limited only by the mesa boundary. We have recently studied the behavior of indirect excitons in such an open geometry, and found an abrupt phase transition at a critical temperature and excitation power density [12]. The PL separates into two spatial regions, one which consists of electron-hole plasma and another that has a set of properties of a high density liquid.

In this work we investigate this phase transition using spatially resolved PL and resonant Rayleigh scattering (RRS). Measuring the threshold power density as a function of temperature we determine the phase diagram of the system over the temperature range 0.1 – 4.8K. Pump probe measurements, in which the liquid is created by a focused pump beam and studied by a much weaker probe, reveal that the liquid is dark and diffuses to large distances away from the illuminated spot, filling the entire area of the mesa at low temperatures. We find that the RRS spectrum narrows significantly and becomes uniform over macroscopic distances at the liquid phase, indicating that the disorder in the sample is effectively screened.

The sample structure is identical to that used in [12] and consists of two GaAs quantum wells with well widths of 12 and 18 nm, separated by a 3-nm $Al_{0.28}Ga_{0.72}As$ barrier. Top and bottom n-doped layers allow the application of voltage that shifts the energy levels of the wide well (WW) relative to the narrow well (NW) [15]. To create the liquid we apply voltages exceeding -2.4V and excite the system with a laser diode at energy of 1.590eV, focused to a Gaussian spot with 22μm half width at half maximum (HWHM) [16].

Figure 1 shows the evolution of the PL spectrum with power at T=0.3K and V=-3.0V. At power density levels below threshold the PL spectrum is dominated by the WW direct exciton peak, $X_{WW}$, at 1.524eV [16]. The lower panel shows the corresponding image of the spatial distribution of the total PL intensity from the mesa: A Gaussian profile, reflecting the shape of the illuminating beam, is observed. When the power is increased beyond a threshold power $P_C = 30μW$, corresponding to a power density of $1.3W/cm^2$, we observe a sudden change in the PL spectrum and spatial distribution: a new broad spectral line, Z, appears at 1.518eV, and a dark ring forms around the center of the beam. This spectral line was identified in [12] as due to indirect recombination of electrons and holes in a liquid state, and the dark ring is formed due to the repulsion of the excitons in the gas by the high-density dipoles at the liquid [12]. As the power is further increased the dark ring shrinks in size and eventually disappears. This behavior is robust and remains qualitatively the same as we change the spot size, such that the threshold power density remains at ~ 1W/cm$^2$. In the measurements that are described throughout the paper the spot size is kept constant with 22μm HWHM.

Spatially resolved measurements, where a pinhole of 10μm diameter is used to resolve the PL, reveal that the Z line appears only at the outer region that surrounds the dark ring, while the $X_{WW}$ line appears in the inner region. Hence, the differences in the relative strength of the $X_{WW}$ and Z peaks in Fig. 1 reflect simply the changes in the relative areas of the inner and outer regions.

We determine the gas-liquid phase diagram by measuring the threshold power $P_C$ as we vary the temperature at constant voltage (blue dots Fig. 2). The direct proportionality between power and carrier density in the gas phase, where recombination is dominated by Xww, implies that we are in fact determining the phase diagram in the carrier density – temperature plane. It is seen that the plane is divided into two regions, a first region of gas only and a second region where liquid appears, with a line of coexistence between them. It is instructive to compare this coexistence line to the Clausius-Clapeyron (C-C) relation, $d\mathcal{P}/dT = L/(T\Delta v)$, which describes a general gas – liquid phase transition [17]. Here $\mathcal{P}$ is the gas pressure, L is the latent heat and $\Delta v$ – the specific volume change in the transition. In the dipolar exciton gas, the pressure can be expressed as $\mathcal{P} = \alpha n^2$, where $n$ is the exciton gas density and $\alpha = e^2 d/\varepsilon$ ($d$ is the distance between the center of the wells, and ε is the dielectric constant) [18], and $\Delta v$ can be approximated by $n^{-1}$. Hence, the C-C equation takes the form $dn/dT = L/(2\alpha T)$. The inset of Fig. 2 shows the resulting $L(T)$ obtained from the experimental data by taking the value of $T\Delta P_C/\Delta T$ at each temperature (in units of $L_0$ defined below). Two regimes can be clearly identified: (i) $T \lesssim 1.1$K where the latent heat exhibits a linear dependence on T, extrapolating to

$L \approx 0$ at $T = 0K$; (ii) $1.1 \lesssim T < 4.8K$ where the latent heat is constant and equals $L_0$. (The transition temperature between the two regimes is in the vicinity of 1.1K but was not determined precisely.) With these assumptions the C-C equation could be easily integrated, and fitted to the experimental data (solid lines in Fig. 2). We confirm the validity of this phase diagram by repeating this measurement under different conditions, where the pump energy is tuned to 1.530eV, such that carriers are excited in the WW only. A similar phase diagram, with lower threshold power, is obtained. Remarkably, the latent heat exhibits exactly the same behavior with two temperature regimes [16].

We note that a linear dependence of $L$ on temperature is expected for a Bose-Einstein condensation (BEC) transition [19]. Indeed, this is the predicted behavior of dipolar excitons at low temperatures [20]. Hence, our findings could be considered as an indication that the exciton liquid at low temperatures undergoes a BEC. We can estimate the value of $L_0$ using the measured critical density of the gas at 1.5K, ~$2 \times 10^{10}$ cm$^{-2}$, and $d = 18$nm [12]. The resulting latent heat per particle is $L_0 \approx 1$meV, consistent with the interaction energy scale of the dipolar exciton liquid, and similar to that of $^4$He [21].

We turn now to describe our RRS measurements. It is well known that RRS is an insightful probe of critical phenomena because of its sensitivity to the disorder in the sample, which often changes near a critical point [22, 23]. We perform our measurements in a pump-probe configuration, creating the liquid by the laser diode pump and measuring the scattered intensity of a much weaker Ti:Sapphire probe, tuned to the NW exciton resonance. We use the fact that the RRS signal should be strongly enhanced at an exciton resonance, and its linewidth is a measure of the disorder in the sample [24, 25]. Hence, the RRS of the NW direct exciton can serve as an effective probe for the changes induced by the liquid.

Firstly we study the behavior at low power density and voltage, when tunneling from the WW to the NW is inhibited. The RRS signal is easily detected in this range as a sharp spectral line at the laser frequency, riding on the broad NW PL spectrum, and its lineshape is obtained by tuning the laser frequency and measuring the scattered intensity. A typical spectrum is depicted by the dashed line in Fig. 3(a). Its spectral width, Γ=0.3 meV, remains unchanged in the relevant temperature range, $0.1 - 5K$, and manifests the static disorder in the sample due to well width and alloy fluctuations. As we increase the voltage the NW is populated by electrons that tunnel from the WW, the NW luminescence becomes trionic, and the RRS signal gradually diminishes, such that above -3.0 V it is undetectable [16].

This behavior changes as we cross the transition threshold, and a strong RRS signal abruptly appears. The open circles in Fig. 3(a) show the RRS spectrum at T=0.6K when the pump power is set to be above $P_C$. We find that the spectrum becomes significantly narrower, $\Gamma$=0.12 meV, and exhibits a strong temperature dependence (Fig. 3(c)), in contrast to the behavior below threshold. The two temperature regimes identified in the phase diagram (Fig. 2) are well manifested in the temperature dependence of the RRS spectrum. We find that at $T \lesssim 1.1$K the RRS spectrum remains unchanged as we vary the temperature, keeping the same width and height, and it rapidly broadens and weakens at $1.1 < T < 4.8$K. As we approach $T_C$ =4.8K [26] the RRS signal drops sharply, and cannot be observed at any power or voltage above that temperature (Fig. 3(c) and (d)).

The solid line in Fig. 3(a) is a fit of the measured points at 0.6K to a Lorentzian: A good agreement is obtained, suggesting that the excitonic line is predominantly homogeneously broadened at $T \lesssim 1.1$K [16]. Since the RRS signal is collected from the entire probe area, this narrow and symmetric line implies that the static disorder potential in the NW is *effectively screened* at the liquid phase. This conclusion is corroborated by spatially resolved measurements. Figure 3(b) shows an image of the mesa emission when the probe frequency is set at the NW exciton peak. It is seen that the RRS signal comes mainly from bright spots at fixed locations on the mesa. We find that while the intensity of the spots varies, their spectra are rather uniform. Analysis of the scattering from spots that are located tens of μm away from one another reveals that their spectral widths are similar and the variation between their peak energy is smaller than the laser spectral resolution, 20μeV [16]. Such a uniform RRS spectral signal of remote locations implies that the energy landscape seen by the NW excitons is *nearly flat over macroscopic distances*.

Screening of the disorder potential by dipolar exciton liquid was predicted and calculated in several theoretical works [27, 28]. It was shown that the repulsive dipole-dipole interaction pushes the excitons to the local potential minima, and thereby changes their density profile. This generates a smoother potential that is the sum of the random potential and the Hartree repulsion by the localized excitons. In the quantum degenerate regime this screening is expected to be strongly enhanced, such that a disorder potential of 0.5 meV amplitude should be practically screened out for densities exceeding $10^{10}$cm$^{-2}$ [27]. Our findings are to our knowledge the first direct observation of this screening effect.

To obtain further insight into this screening effect we study the spatial dependence of the RRS intensity. We find that at intermediate pump power levels, P~40μW, the RRS signal exhibits spatial phase separation, similarly to the PL signal. This is demonstrated in Fig. 4(a), where the stark contrast between the two regions and the dark phase boundary are clearly observed. It is seen that strong RRS is obtained only at the outer parts of the illuminated area, confirming the global nature of the liquid formation. We then shift the probe beam to the side of the mesa and examine the RRS signal away from the pump beam (Fig. 4(b)). Remarkably we find that the RRS signal in this area is similarly intense and exhibits the same behavior as in the main beam: Abrupt enhancement of intensity as the pump power crosses threshold, narrow linewidth and a spotty pattern. This behavior is found as we position the probe beam *anywhere in the mesa*, proving that the liquid diffuses beyond the pump region into the whole area of the mesa.

We find that the liquid diffusion *decreases* with increasing temperature: While at 0.6K the RRS intensity (averaged over the area of the probe) remains constant as we shift the probe throughout the mesa, at 4K it drops by a factor of five between overlapping (Fig. 4(a)) and non-overlapping (Fig. 4(b)) configurations. This unique temperature dependence is a manifestation of the repulsive interaction, $E_{int}$, which characterizes the dipolar exciton liquid, and enhances the diffusion constant by a large factor, $E_{int}/T$ [27, 29]. In the quantum regime this diffusion constant is expected to be further increased, by another factor of $\exp(T_C/T)$ [27], allowing the excitons to cross macroscopic distances within the liquid lifetime [12].

In the concluding part we wish to return to the PL measurements, this time performed in a pump-probe configuration, where we set the pump and probe energy at 1.590 eV, and collect the probe PL from the area marked by the small rectangle in Fig. 5(a). We find that the PL spectrum of the probe changes as we switch on the pump and set its power to be above $P_C$, in a similar manner to the changes observed for the pump PL (Fig. 5(b)). Shifting the probe beam around reveals that the liquid spectrum is observed throughout the mesa, yet we could detect light emission only when the liquid is probed by an exciting beam. This is demonstrated in Fig. 5(c), which shows the Z line intensity as a function of probe power. It is seen that the PL intensity extrapolates to approximately zero when the probe power vanishes. We conclude that the Z line is not due to recombination of the liquid carriers, as we postulated in [12], but rather - recombination *in the presence* of the liquid. It is plausible to assume that it is due to recombination of carriers in the NW and WW which have not relaxed into the condensate with opposite charge carriers from the condensate.

Evidence for the non-radiative nature of the dipolar exciton liquid was recently reported in [11]. It was suggested that the excitons condense in a "dark" state, consisting of electrons and holes in parallel spin configuration that cannot couple to light. This state is lower in energy than that of the bright exciton, and should therefore be the ground state for BEC [30]. One would expect, however, scattering mechanisms to give rise to some emission, turning the condensate to become "gray" [31]. We find no clear evidence for such emission within our detection limit, set by the Gaussian tail of the pump [16]. Further experiments are needed to provide explanation for the origin of this effect.

Acknowledgments: This work was supported by the Israeli Science Foundation.

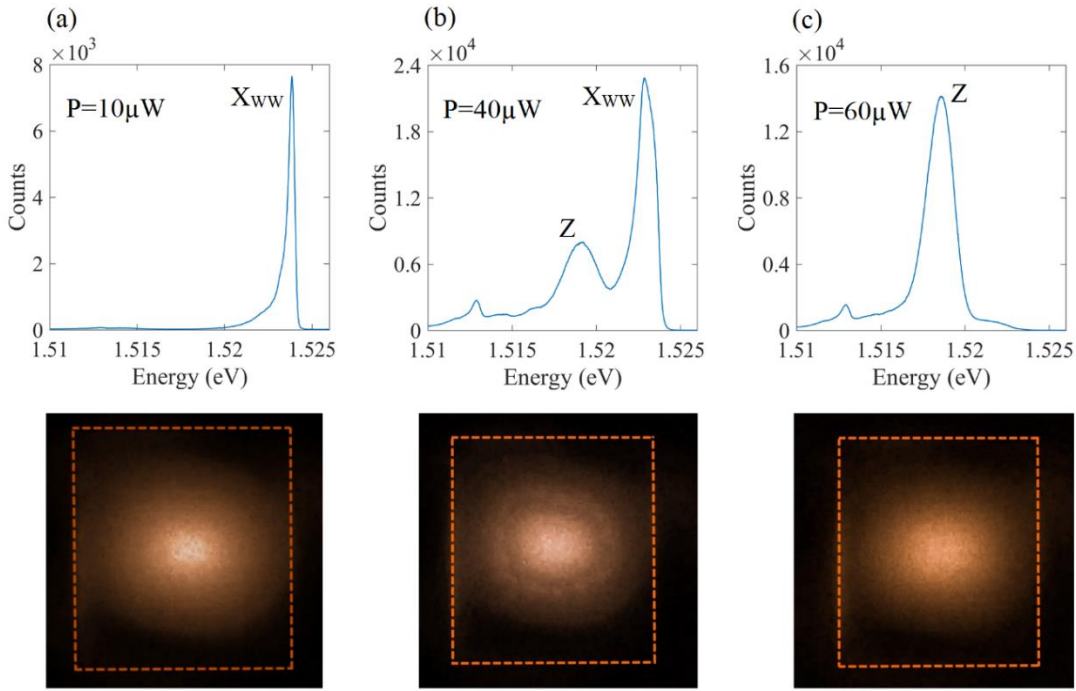

Figure 1: The upper and lower panels show the PL spectra and PL images, respectively, at different pump powers at a fixed $V_g=-3.0V$ and T=0.6K. The mesa is 100μm×100μm and is marked by dashed lines. (a) At low power, P=10μW, the WW direct exciton peak, $X_{WW}$, is observed and the image is a smooth Gaussian (the low energy shoulder is due to a trion peak). (b) At power above threshold, P=40μW, a new broad spectral line, Z is observed in the spectrum and correspondingly, a dark ring appears in the PL image around the center of the illumination spot. (c) As the power is further increased, P=60μW, the $X_{WW}$ line disappears and the emission spectrum is dominated by the Z line. Also, the dark boundary disappears in the PL image.

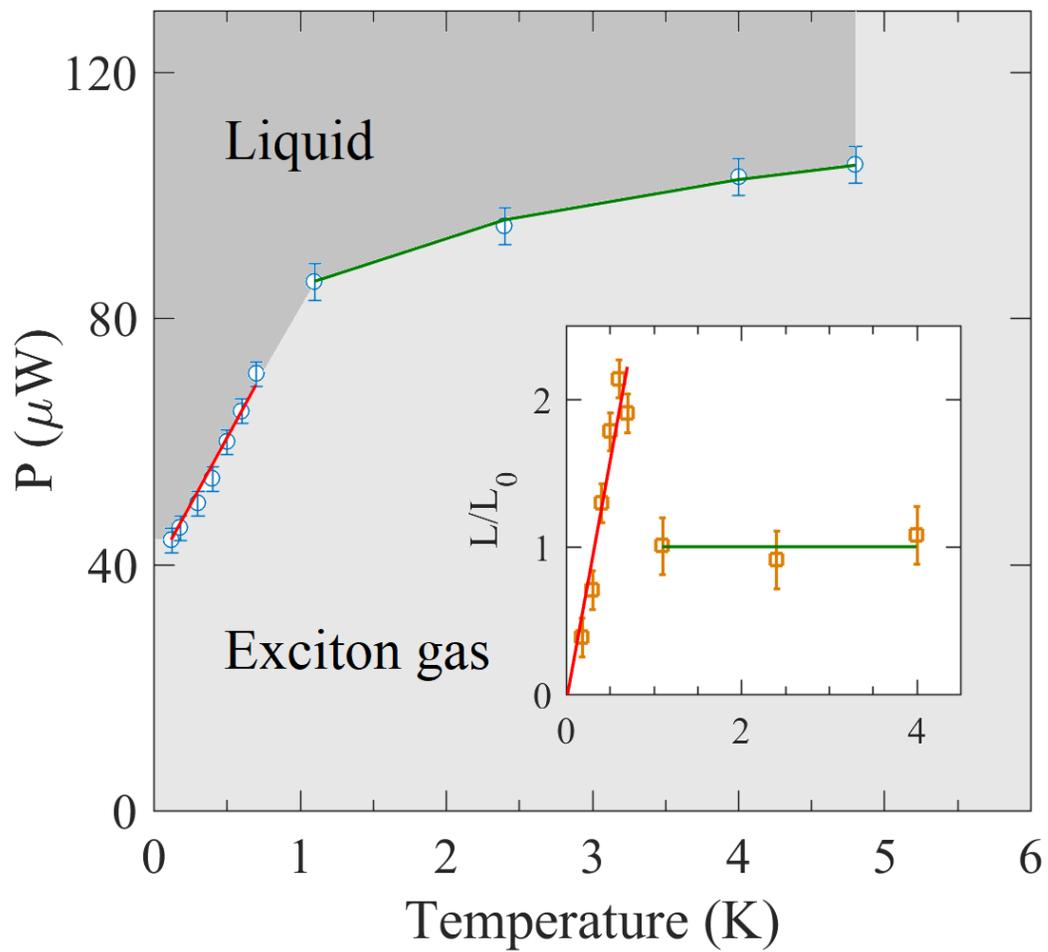

Figure 2: The phase diagram of the system obtained by measuring the threshold power $P_C$ as a function of temperature at V=-2.6V. The inset shows the latent heat $L$ as a function of temperature using the Clausius-Clapeyron (C-C) equation. The solid lines in the main figure are fit to the integrated C-C equation using the values of $L$ from the inset.

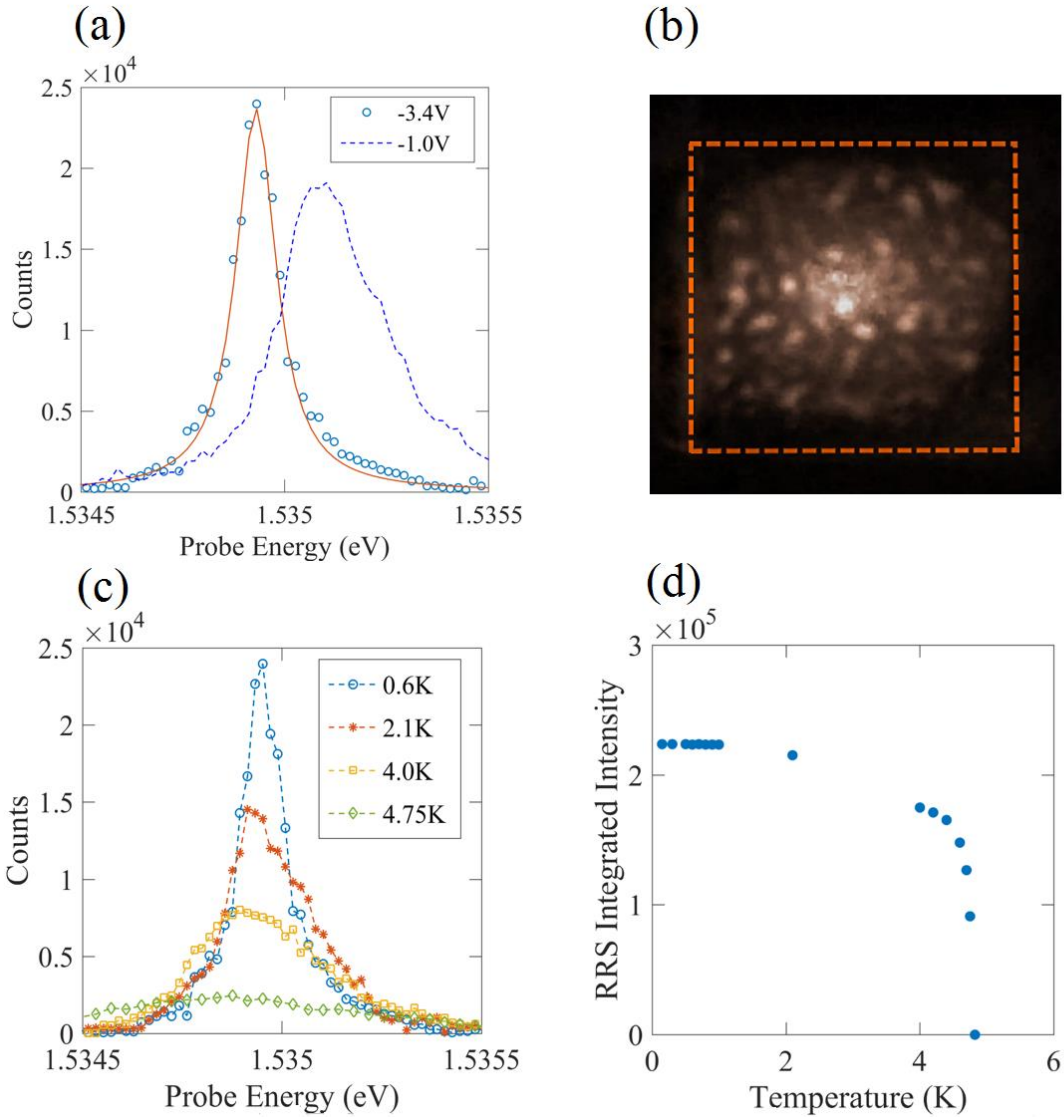

Figure 3: RRS measurements with overlapping pump (50μW) and probe (5μW) beams at V=-3.4V. (a) The measured RRS spectrum at T=0.6K (blue circles) and a Lorentzian fit (solid red line). The dashed line shows the RRS spectrum at pre-liquid voltage, V=-1V. (b) Image of the mesa emission (RRS + PL) when the probe energy is set at the NW exciton peak and T=0.6K. The spotty RRS pattern is clearly observed. (c) The RRS spectra at several temperatures below $T_C$. (d) The RRS integrated intensity as a function of temperature. It is seen that it is constant at $T < 1.1$K, then changes by less than 20% at $1.1 < T < 4$K, and rapidly diminishes towards $T_C$.

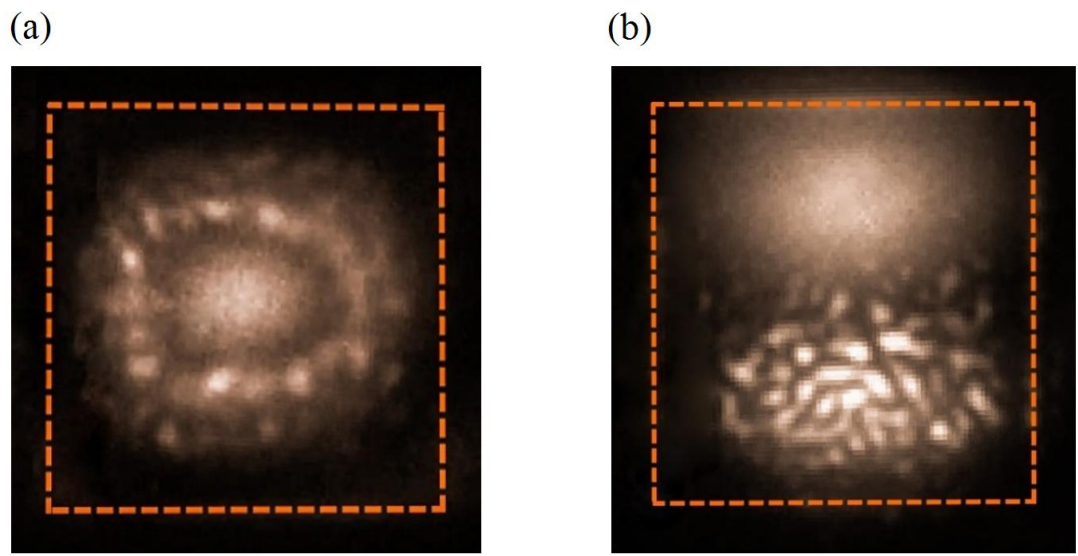

Figure 4: Images of the mesa emission (RRS + PL) for probe excitation in resonance with the NW exciton at -3.0V and T=0.6K. (a) Overlapping pump and probe at intermediate pump power (40μW), where phase separation occurs. (b) Non-overlapping pump (at the top) and probe (at the bottom) at high pump power (60μW). The orange dashed square marks the mesa.

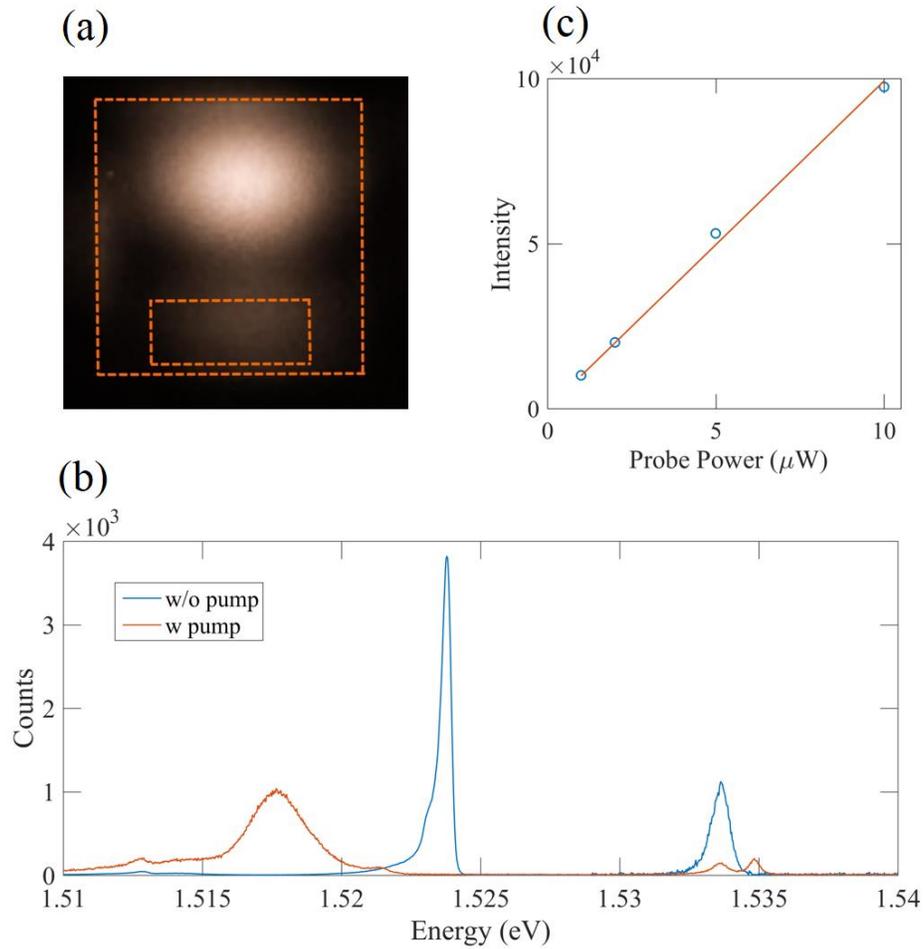

Figure 5: Two-beam PL measurements at P=50µW, V=-3.0V, T=0.6K. (a) A typical setting of the pump (top) and probe (bottom) experiment. The small rectangle marks the area from which PL is collected. (b) The PL spectrum from the WW and NW without (blue) and with (red) the pump beam. The indirect exciton appears as a weak peak at 1.500eV, outside the range of this figure [16]. (c) The integrated PL intensity from the marked area as a function of probe power when the pump is switched on.

# Experimental Study of the Exciton Gas – Liquid Transition in Coupled Quantum Wells

## Supplementary Material

Subhradeep Misra[1], Michael Stern[2], Arjun Joshua[1], Vladimir Umansky[1] and Israel Bar-Joseph[1]

[1]*Department of Condensed Matter physics, Weizmann Institute of Science, Rehovot, Israel*
[2]*Institute of Nanotechnology and Advanced Materials, Bar-Ilan University, Ramat-Gan, Israel*

# Contents

# Materials & Methods

## (a) Sample Structure

A schematic description of the sample and its constituting layers can be seen in Fig. S1(a). The coupled quantum wells (CQW) system is the same as in [1] and consists of two GaAs quantum wells having widths of 12 and 18 nm that are separated by a 3 nm wide $Al_{0.28}Ga_{0.72}As$ barrier. It is embedded in a 2 μm wide n-i-n structure so that a voltage can be applied in the perpendicular direction. The top and bottom $n^+$ layers are silicon doped ($n_{Si}$~$10^{18}$ cm$^{-3}$) $Al_{0.12}Ga_{0.88}As$. The intrinsic region consists of two superlattice (SL) layers, each 1 μm thick: below the CQW, we have 33 periods of [27 nm $Al_{0.37}Ga_{0.63}As$] − [2nm AlAs] − [1 nm GaAs], and above the CQW, 20 periods of [50 nm $Al_{0.33}Ga_{0.67}As$] − [1nm GaAs]. This SL structure ensures that the crystal quality is maintained, as indeed reflected in the PL linewidth of the two wells, which is sub-meV. The band gap of the contact layers, 1.650 eV, and the SL layers, ~1.9 eV, is selected to be well above the illumination energy of the laser so that these layers do not contribute to carrier photo-generation. This eliminates the photo-depletion mechanism that gives rise to the ring formation [2]. The sample was processed using optical lithography, wet etching and metal deposition by thermal evaporation technique to form square mesas, each having 100μm sides, resulting in an area 0.01 mm$^2$. The experiments reported in this paper were conducted on several mesas and were found to be highly reproducible.

Upon application of a high enough negative voltage (<-1 V) to the top layer, the electron level in the wide well (WW) becomes higher than the corresponding level in the narrow well (NW) and electrons can tunnel from the WW to the NW (Fig. S1(b)). The tunneling rate of the holes from the NW to the WW is much lower, and grows as the magnitude of the voltage increases.

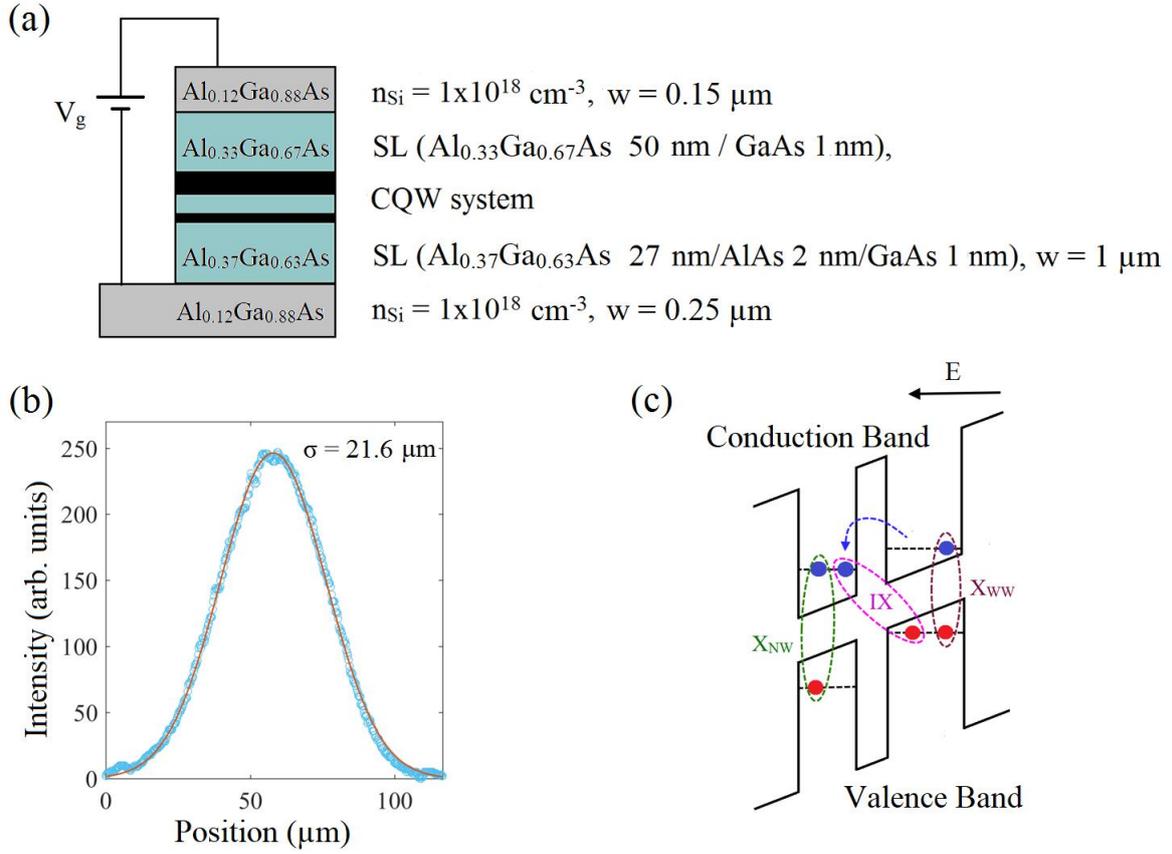

Figure S1: *(a) A schematic diagram of the sample. The CQW consists of two quantum wells, 18 and 12 nm wide, with a 3 nm $Al_{0.28}Ga_{0.72}As$ barrier. (b) The illumination beam profile fitted to a Gaussian function. (c) The energy band diagram of the asymmetric CQW system under the influence of an applied electric field in the growth axis. Electrons and holes are represented by blue and red circles, respectively. Under this band alignment, for illumination energy above both WW and NW, excitons are created in the two wells, $X_{WW}$ and $X_{NW}$. Tunneling of an electron can occur from WW into the NW conduction band (blue dashed arrow), and by binding weakly to a hole in the WW form an indirect exciton, IX.*

## (b) Experimental Setup

Experiments are conducted in a dilution refrigerator with optical windows, having a base temperature of 100 mK. A careful design of the temperature gradient in helium reservoir surrounding the vacuum chamber eliminates the formation of helium bubbles and the laser scattering associated with them. The high numerical aperture of the cryostat optical windows, 0.4, allows effective collection of the PL. We have conducted the measurements over a temperature range of 0.1 - 6 K.

One notable fact here is that upon various instances of thermal cycling between 4K and room temperature, the electrical properties of the sample deteriorated, and its leakage current increased. However, no changes were noticed in the optical properties, implying no appreciable degradation in quality.

For most of the experiments the sample is illuminated by a laser diode with a Gaussian spot at energy of 1.590 eV (Fig. S1(b)). The spot has a half width at half maximum of σ = 22 μm. For the two-beam experiments we shifted the beam to the side of the mesa, such that the further parts of the mesa are illuminated by less than 1% of the power density at the peak. The probe laser is a tunable Ti:S laser, having the same spot-size as the pump beam, and its power is adjusted such that pump-probe power ratio remains 10. The signal is analyzed using a U-1000 Jobin-Yvon Raman spectrometer and imaged on a CCD camera (Prosilica GC1380) with an overall spatial resolution of 4μm.

## 1. Photoluminescence Measurements
### (a) Low power measurements

Figure S2 shows the low temperature (T = 0.3 K) PL spectrum of the system for very low excitation power (P = 50 nW) as a function of applied voltage (The laser excitation energy is 1.590 eV). The direct exciton PL from the WW ($X_{WW}$) is seen at 1.524 eV, and 1 meV below it we observe the WW trion line (The broad peak at 1.521 eV is an impurity line that disappears at higher excitation power). The indirect exciton (IX) peak is easily recognized: It crosses the $X_{WW}$ line at -1 V and its energy falls linearly for further increment of the negative voltage. At -3 V the IX energy is 1.5 eV, approximately 25 meV below the $X_{WW}$ line. We note that at the crossing voltage the energy drop between the centers of the WW and the NW is 9 meV. Indeed, as evidenced by the PL spectrum this is the required energy drop to bring the electron level in the NW to be below the corresponding one in the WW.

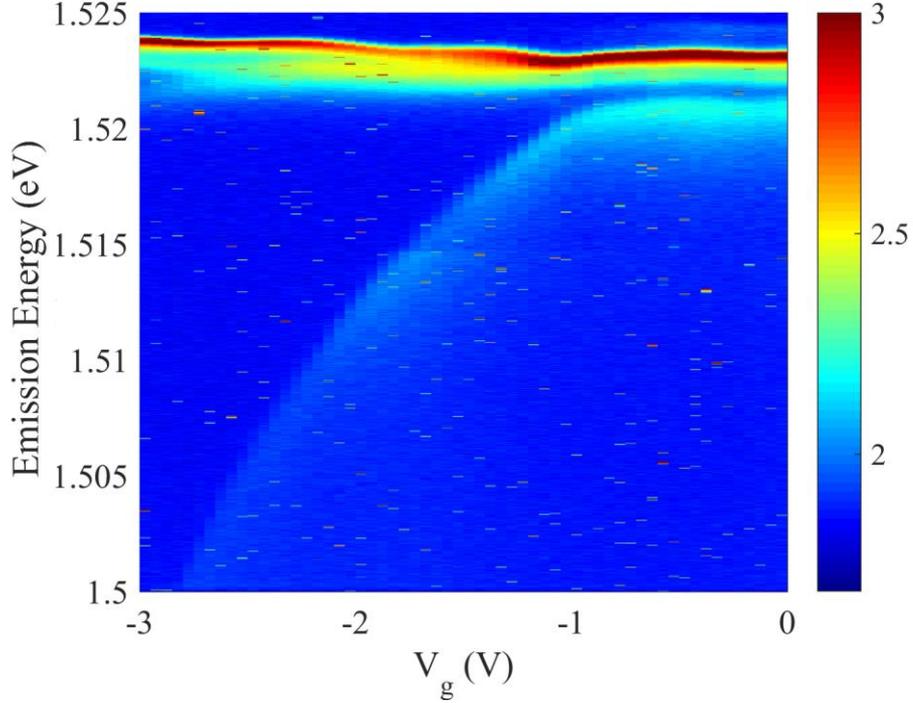

Figure S2: *The voltage ($V_g$) dependence of the PL spectra for very low illumination power, P=50 nW. The PL intensity is color coded and presented on a logarithmic scale. The excitation energy is 1.59 eV and T=0.3 K. The $X_{WW}$ energy remains nearly constant at 1.524 eV in the whole range, V=0 to -3.0 V. For voltages more negative than -1.0 V, the linear V dependence of the IX energy is clearly observed.*

### (b) The Photoluminescence spectrum of the narrow well exciton

Figure S3 shows the probe PL spectrum at the energy range of the NW as a function of gate voltage, at pump power of 50 µW (The PL intensity is color coded on a logarithmic scale). At low negative voltages, $V_g > -1$ V, when tunneling from the WW to the NW is inhibited, the system PL spectrum is excitonic. As we increase the voltage magnitude beyond -1 V tunneling is allowed and the NW is filled by electrons. This is reflected in the PL spectrum which becomes dominated by the negatively charged trion peak 1 meV below. At -2.6 V we cross the threshold for the gas liquid phase transition and the direct and trionic transitions are strongly suppressed, as seen in Fig. 3(b) of the main text.

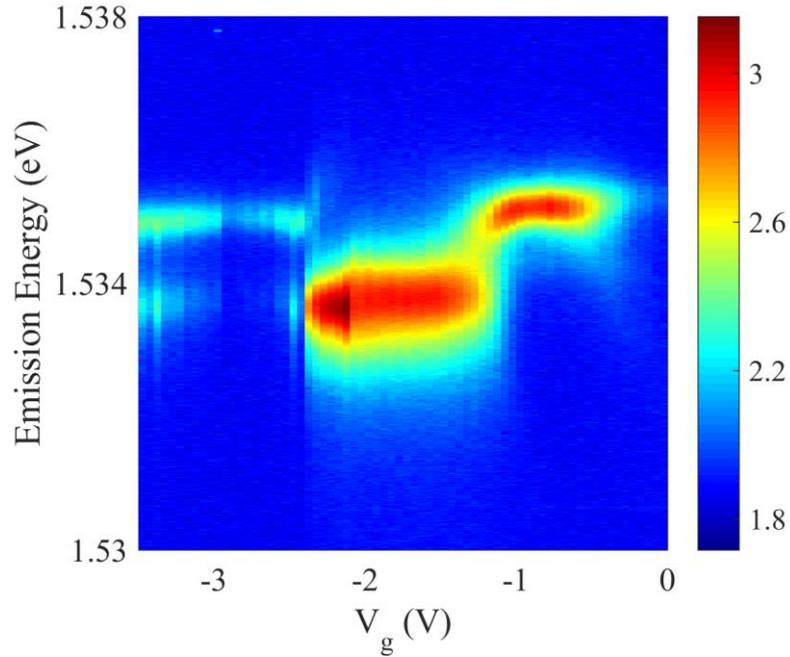

Figure S3: *The probe PL spectrum of the NW as a function of voltage. The PL intensity is color coded and presented in a logarithmic scale. The measurements are done at T=0.5 K, with excitation energy 1.59 eV and pump and probe powers being 50 µW and 5 µW respectively.*

## 2. Properties of the liquid
### (a) Voltage Dependence

The gas-liquid phase diagram of the system in the power-temperature plane for temperature range 0.1-4.8 K is discussed in the main text. This diagram is obtained by measuring the threshold power, $P_C$ at various temperatures. One important aspect of the phase transition is that for a fixed temperature, as we set the voltage to more negative values, $P_C$ becomes lower, changing by approximately a factor of two between -2.6 V and – 3.0 V. This is demonstrated in Fig. S4(a).

This result can be interpreted in the following way: The carrier density is directly proportional to the radiative lifetime ($\tau$) of the carriers, $n = P\tau$, where $P$ is the power density. When the applied electric field is increased, the carriers become more localized by triangular wells which in turn increases their wave-function confinement. This reduces the electron-hole overlap integral and thus increases the radiative lifetime. As the voltage is set to more negative values, less excitation power is required to obtain the critical carrier density, which results in lowering of $P_C$. Indeed, the behavior of the PL described in Fig S4 is reproducible when the voltage is varied instead of the excitation power, which further validates this interpretation.

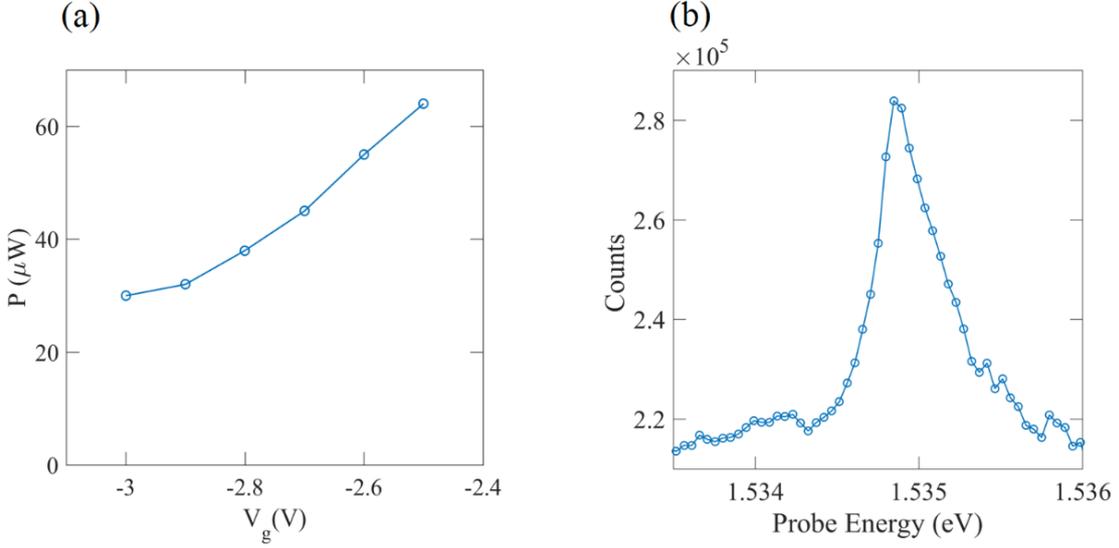

Fig. S4: *(a) The voltage dependence of the liquid phase, obtained by measuring the threshold power, $P_C$, as a function of $V_g$ keeping the system temperature at 0.5 K. (b) The liquid luminescence obtained from the probe beam using a 10 µm diameter pinhole in a two beam experiment. The liquid is created with a strong pump beam, P=50 µW at $V_g$ = -3.4 V and is probed by a spatially separated weak beam (P = 5 µW) tuned near the NW exciton resonance. The enhancement of the liquid luminescence at the resonance confirms the role of the NW carriers in the formation of the liquid.*

### (b) The Indirect Nature of the Liquid Line

In this section we prove that the Z line is due to indirect recombination of electrons from the NW with WW holes. We first notice that the Z line can be created with WW excitation only, when the laser is set to 1.530 eV (See next section for further details). This proves that WW carriers are involved. We then have to establish that NW carriers are also involved. This is achieved by pumping the system at the energy range near the NW exciton and collecting the intensity of the Z line, essentially performing a PL excitation experiment (PLE). We find that the PL intensity is enhanced as the excitation energy is tuned across the NW exciton resonance, following its absorption spectrum (Fig. S4(b)). The PLE intensity is peaked at 1.535 eV, coinciding with the NW exciton PL. This proves that the NW electrons are an integral component of the liquid.

## 3. The Gas Liquid Phase Diagram

To confirm the general nature of the phase diagram we measured it under very different excitation conditions, where the pump energy is set at 1.530 eV, 4 meV below the NW exciton energy, which is at 1.534 eV. Under these excitation conditions electron-hole pairs are generated in the WW only, and above a certain voltage – electrons can tunnel to the NW. To determine the phase diagram we performed a measurement of the threshold power as a function of temperature at -2.6 V. Figure S5 compares the phase diagram obtained when the system is excited at the NW (a) and the WW (b), at the same voltage. We can see that while the threshold values are ~ 30% lower in the second case, we get the same dependence as a function of temperature: a linear increase at low T, followed by a slower, logarithmic increase at higher temperature. The insets show analysis of the latent heat as a function of temperature, using $L \propto T \Delta P_C / \Delta T$, where $\Delta P_C$ is obtained from the experimental data by averaging the difference in threshold power in the range $T - \Delta T$ to $T + \Delta T$. Remarkably, the two measurements yield the same behavior for $L/L_0$.

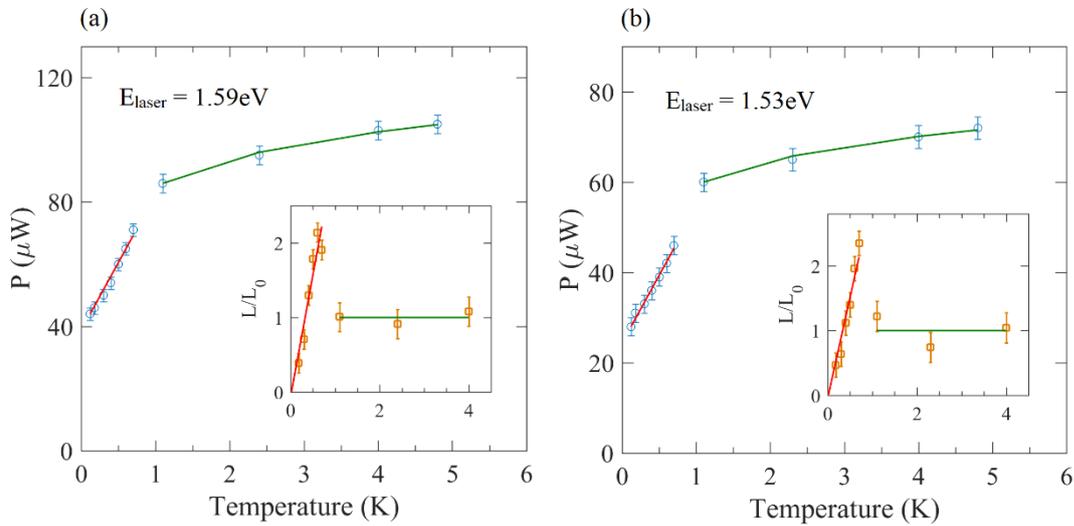

Fig. S5: *The threshold power as a function of temperature at V = -2.6 V when the system is excited at (a)1.59 eV and (b) 1.53 eV. The inset shows the latent heat L as a function of temperature. The solid lines in the main figure are fit to the integrated C-C equation using the values of L from the inset.*

# 4. Resonant Rayleigh Scattering (RRS) Measurements
## (a) The Resonant Rayleigh Scattering Signal

Figure S6 shows the measured RRS signal at the spectrometer as the laser energy is tuned over 1 meV range, between 1.5345 eV to 1.5355 eV, at $V_g$ = -3.4 V with illumination pump beam power 70 µW and at temperature of 0.6 K. To emphasize the relative strength of the PL and the RRS signal we present in Fig. S6(a) a color plot of the signal intensity using a logarithmic scale. The scattered laser signal appears as a straight line in the plane, where the detected energy equals the exciting laser energy, while the PL is seen as a broad weak tail below the excitation energy. The measured RRS spectra for different excitation energies across the NW exciton resonance are shown in Fig. S6(b). It can be seen that the scattered light in the off-resonant excitation is very weak (peak intensity ~ 150 counts) and it enhances significantly (~ 100 times) at the resonance. The inset shows the measured spectrum slightly above the exciton peak: the PL signal is clearly observed as a low energy tail.

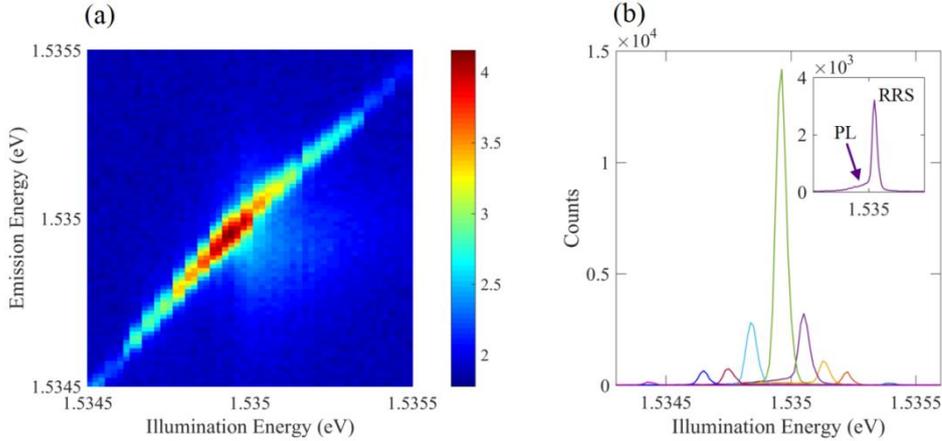

Figure S6: *The RRS measurements in the two beam experiment, with the signal obtained from the probe beam, $V_g$ = -3.4 V, P = 70 µW, T = 0.6 K. (a) The emission spectrum as a function of laser illumination energy. The intensity is color coded and presented on a logarithmic scale. The enhancement in the vicinity of the narrow well excitonic resonance is clearly seen. The weak PL from the excitonic recombination is also observed as a low energy weak signal. (b) Spectra for different illumination energies above and below the NW resonance. The large enhancement of the scattered light at the resonance as compared to the off-resonant illumination is clearly demonstrated. Inset: The weak PL can be distinguished from the strong RRS signal for illumination slightly above the resonance peak.*

We find the RRS signal is partially polarized, with approximately 4% of the signal at the crossed polarization. The reported measurements are performed at a polarization angle slightly off the crossed polarization angle.

### (b) The Resonant Rayleigh Scattering spectrum of the various spots

We find that the RRS signal exhibits a spotty pattern, both at the pre-liquid and liquid conditions. These spots are not speckles, but rather regions in the sample where the scattering cross section is large. We verified this by changing the imaging setup and observing that the image remained unchanged for different magnifications. Atomic force microscope (AFM) measurements of the sample gave a very flat surface, with no indication for structural defects that could explain these spots.

Figure S7 shows the RRS spectrum of three spots which are tens of μm away from each other. We obtained these spectra in the following way: We first imaged the RRS signal from the mesa at different laser wavelengths, with 0.01 nm step (corresponding to approximately 20 μeV). We then obtained the RRS spectrum for each spot by plotting the measured scattered intensity as a function of excitation wavelength. We find that while the scattering amplitude of the various spots changed significantly, the variation in the peak energy and width are very small: In 10 points that were analyzed they are equal to or smaller than our spectral resolution, namely 20 μeV.

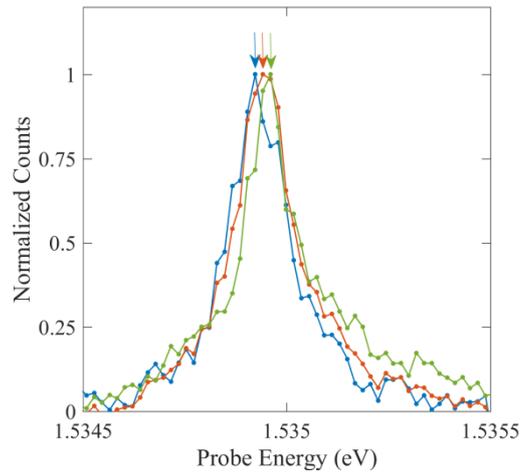

Figure S7: *Normalized spectra of individual bright RRS spots separated tens of μm away from one another measured at T = 0.6 K, V=-3.4 V, P=50 μW. The arrows point to the peak position of the corresponding spectrum. We observe that all spots show qualitatively the same spectral profile as the integrated RRS signal in Fig.3(a) in the main text.*

### (c) The Lorentzian nature of the Resonant Rayleigh Scattering spectrum at low temperature

The RRS spectrum in the temperature range below 1.1 K is constant in shape (width and height) and can be fitted to a Lorentzian (Fig. 3(a)). In the following we substantiate the choice of a Lorentzian over a Gaussian fit. In Fig. S8 we show the measured liquid RRS spectrum of Fig. 3(a) (red circles), presented on a logarithmic scale, together with a fit to a Lorentzian (blue line) and to a Gaussian (brown line). It can clearly be seen that the slow decay of the tail is consistent with a Lorentzian line-shape and cannot be fitted by a Gaussian.

At higher temperatures the RRS spectrum broadens and acquires an asymmetric shape, which clearly deviates from a Lorentzian shape, nevertheless its area remains nearly constant up to 4 K (Fig. 3(d)). This behavior is not clearly understood.

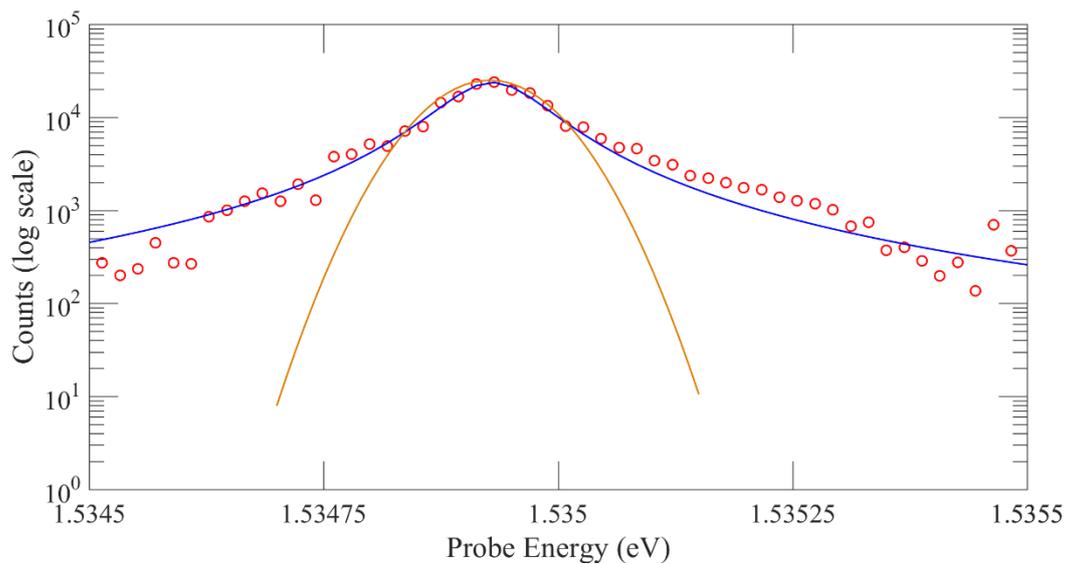

Fig S8: *The measured liquid RRS spectrum of Fig. 3(a) (T = 0.6 K, V = -3.4 V, P = 50 µW), presented on a logarithmic scale. The blue line is a fit to a Lorentzian and the brown – to a Gaussian.*

## 6. Photoluminescence darkening

We find that at threshold the total PL intensity starts to decrease, an effect reported by several previous groups (see for example [3]). This "PL darkening" was interpreted as an indication for condensation in a dark exciton state: the electrons and holes can only recombine non-radiatively and hence there is a decrease in the total PL intensity. Figure S9 demonstrates this effect in a pump-probe experiment. Here the normalized PL intensity of the probe is plotted as a function of the pump power. It is seen that PL darkening of the probe sets in exactly at threshold, indicating that non-radiative recombination starts playing a role. We note, however, that the transition is accompanied by a shift of the PL emission from being primarily due to direct recombination before the transition to indirect after the transition. This by itself would increase the recombination time substantially and give larger weight for non-radiative processes. Hence we do not consider this effect as a direct evidence for condensation.

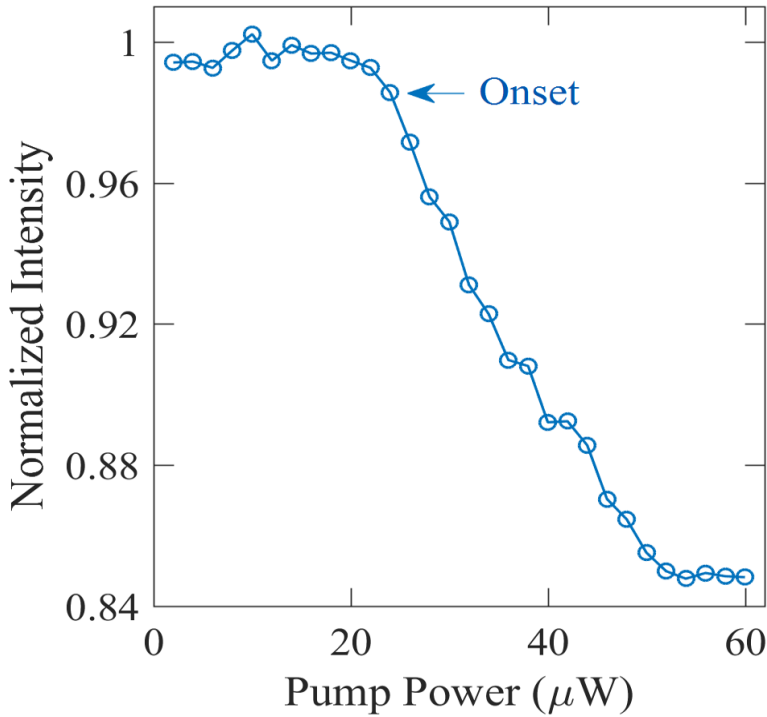

Figure S9: *The evolution of the integrated PL intensity of the probe (see the marked rectangle in Fig. 5(a) in the main text) as a function of the pump power at T = 0.6 K. The probe PL darkening is clearly seen at the onset of the phase transition.*

## 7. The Dark Nature of the Liquid

To establish the dark nature of the liquid we tried to measure the emitted PL in the dark regions of the sample under conditions where the probe is off and only the pump is present. We find that the Z line intensity follows the spatial profile of the pump beam at all temperatures below $T_C$, and there is no evidence for PL emission by the liquid itself.

The upper limit for the intrinsic PL of the liquid can be estimated in the following way: The pump power within a 10x10 μm$^2$ area at the side of the mesa (when the probe is off and the pump is positioned as in Fig. 5(a)) is one to two orders of magnitude (depending on the position of the pump and probe) lower than at the corresponding area near the center of the pump. Hence, assuming that the Z line is due to recombination of an electron in the NW (which isn't part of the liquid) with a hole in the WW (which is part of the liquid), we can conclude that the radiative processes in the liquid are suppressed by at least two orders of magnitude.